\documentclass[12pt,oneside]{article}
\usepackage{graphics,amsmath,amssymb}
\newcommand{\field}[1]{\mathbb{#1}}

\usepackage[left=3.3cm,top=3.5cm,bottom=3.5cm,right=3.3cm,head=0cm,foot=0.7cm]{geometry}

\begin{document}
\title{Explaining Thermodynamic-Like Behaviour In Terms of Epsilon-Ergodicity}

\author{Roman Frigg and Charlotte Werndl\footnote{Authors are listed alphabetically. This work is fully collaborative.} \\
Department of Philosophy, Logic and Scientific Method\\ London School of Economics\\Houghton Street, London WC2A 2AE\\
r.p.frigg@lse.ac.uk and c.s.werndl@lse.ac.uk\\
}

\date{\small{This is a pre-copyedited, author-produced PDF of an article accepted for publication in Philosophy of Science following peer review. The definitive publisher-authenticated version ``Frigg, R. and Werndl, C. (2011), Explaining Thermodynamic-Like Behaviour in Terms of Epsilon-Ergodicity, Philosophy of Science 78 (4), 628-652'' is available online at: http://www.jstor.org/discover/10.1086/661567?uid=3738032\&uid=2\&uid=4\&sid=21102740254323
}}
\maketitle

\abstract{Why do gases reach equilibrium when left to themselves? The canonical answer, originally proffered by Boltzmann, is that the systems have to be ergodic. This answer is now widely regarded as flawed. We argue that some of the main objections, in particular, arguments based on the KAM-theorem and the Markus-Meyer theorem, are beside the point. We then argue that something close to Boltzmann's proposal is true: gases behave thermodynamic-like if they are epsilon-ergodic, i.e., ergodic on the phase space except for a small region of measure epsilon. This answer is promising because there is evidence that relevant systems are epsilon-ergodic.}

\vspace{1cm}

\section{Introduction}

Consider a gas confined to the left half of a container. When the dividing wall is removed, the gas approaches equilibrium by spreading uniformly over the available space. According to the Second Law of Thermodynamics, this approach is uniform and irreversible in the sense that once the wall is removed, the entropy of the system increases monotonically until it reaches its maximum, which it will thereafter never leave. Statistical mechanics (SM) is the study of the connection between micro-physics and macro-physics: it aims to explain the manifest macroscopic behaviour of systems in terms of the dynamics of their micro-constituents.

Such explanations are usually given within one of two theoretical frameworks: Boltzmannian and Gibbsian SM. In this paper we set aside Gibbsian SM and focus on Boltzmannian SM, and we assume systems to be classical.\footnote{For a discussion of Gibbsian SM, see Frigg (2008) and Uffink (2007). For details about quantum SM, see Emch and Liu (2002).} Furthermore, we restrict our attention to gases. These are prime examples of systems in SM, and an explanation of the behaviour of liquids and solids may well differ from one in gases.

After introducing the formalism of Boltzmannian SM (Section 2), we discuss what exactly thermodynamic-like behaviour amounts to (Section 3). Then we review the original ergodic programme and state our own proposal based on epsilon-ergodicity (Section 4). There follows a detailed discussion of the two main `no-go' theorems: the KAM-theorem and the Markus-Meyer theorem. We show that, first appearances notwithstanding, these theorems pose no threat to the ergodic programme (Sections 5 and 6). Furthermore, there are good reasons to believe that relevant systems in SM are epsilon-ergodic (Section 7). We end with some remarks about relaxation times and the scope of our explanation (Section 8) and a brief conclusion (Section 9).

\section{Boltzmannian SM In Brief}\label{BSM}

The object of study in Boltzmannian SM is a system consisting of $n$ classical particles with three
degrees of freedom each.\footnote{For a detailed discussion of Boltzmannian SM, see Frigg (2008, 103--21).} The state of such a system is specified by a point $x$ (the \textit{microstate}) in its $6n$-dimensional phase space $\Gamma$, which is endowed with the standard Lebesgue measure $\mu$. Since the energy is conserved, the motion of the system is confined to a $6n-1$ dimensional energy hypersurface $\Gamma_{E}$, where $E$ is the value of the energy of the system. The time evolution of the system is governed by Hamilton's equations, whose solutions are the phase flow $\phi_{t}$ on the energy hypersurface $\Gamma_{E}$; intuitively speaking, $\phi_{t}(x)$ gives the evolution of $x$ after $t$ time steps. The function $s_{x}:\field{R}\rightarrow\Gamma_{E},\,\,s_{x}(t) = \phi_{t}(x)$ is the \textit{solution} originating in $x$. The measure $\mu$ can be restricted to $\Gamma_{E}$ so that if $\mu$ itself is preserved under the dynamics, then its restriction to $\Gamma_{E}$, $\mu_{E}$, is preserved as well. Furthermore, we can normalise the measure such that $\mu_{E}(\Gamma_{E})=1$ (then $\mu_{E}$ is a probability measure on $\Gamma_{E}$).
The triple $(\Gamma_{E},\mu_{E},\phi_{t})$ is a \textit{measure-preserving dynamical system}, meaning that $\phi_{t}:\Gamma_{E}\rightarrow \Gamma_{E}$ ($t\in\field{R}$) are one-to-one measurable mappings such that $\phi_{t+s}=\phi_{t}(\phi_{s})$ for all $t,s\in\field{R}$, $\phi_{t}(x)$ is jointly measurable in $(x,t)$, and $\mu_{E}(R)=\mu_{E}(\phi_{t}(R))$ for all measurable $R\subseteq\Gamma_{E}$ and all $t\in\field{R}$.

The macro-condition of a system is characterised by \textit{macrostates} $M_{i}$, $i=1,\ldots, m$. In Boltzmannian SM macrostates are assumed to supervene on microstates, meaning that a change in the macrostate must be accompanied by a change in the microstate. This determination relation need not be one-to-one; in
fact, many different microstates usually correspond to the same macrostate. So each macrostate has associated with it a \textit{macro-region} $\Gamma_{M_{i}}$, consisting of all $x\in\Gamma_{E}$ for which the system is in $M_{i}$. The $\Gamma_{M_{i}}$ form a partition of $\Gamma_{E}$, meaning that they do not overlap and jointly cover $\Gamma_{E}$. The \textit{Boltzmann entropy} of a \emph{macrostate} $M_{i}$ is defined as $S_{B}(M_{i})\,:=\,k_{B}\log[\mu(\Gamma_{M_{i}})]$, where $k_{B}$ is the Boltzmann constant; the \textit{Boltzmann entropy} of a \emph{system} at time $t$, $S_{B}(t)$, is the entropy of the system's macrostate at $t$: $S_{_{B}}(t):=S_{_{B}}(M_{x(t)})$, where $x(t)$ is the microstate at $t$ and $M_{x(t)}$ is the
macrostate supervening on $x(t)$. Two macrostates are of particular importance: the equilibrium state, $M_{eq}$, and the macrostate at the beginning of the process, $M_{p}$, also referred to as the `past state'. The former has maximum entropy while the latter is, by assumption, a low entropy state.

An important aspect of the Boltzmannian framework is that for gases $\Gamma_{M_{eq}}$ is vastly larger (with respect to $\mu_{E}$) than any other macro-region, a fact also known as the `dominance of the equilibrium macrostate'; in fact, $\Gamma_{E}$ is almost entirely taken up by equilibrium microstates (see, for instance, Goldstein 2001, 45).\footnote{We set aside the problem of degeneracy (Lavis 2005, 255--58).}

\section{Explaining Thermodynamic-Like Behaviour}\label{Explanation}

A naive approach to SM would first associate the Boltzmann entropy with the thermodynamic entropy and then require that the Second Law be derived from the mechanical laws governing the motion of the particles. This is setting the bar too high in two respects. First, it is impossible to require that the entropy increase be monotonic. The relevant systems show Poincar\'{e} recurrence, and such systems cannot possibly exhibit strict irreversible behaviour because sooner or later the system will return arbitrarily close to its initial condition.\footnote{That this may take a very long time to happen is besides the point as far as a justification of the Second Law is concerned.} We agree with Callender (2001) that thermodynamics is an approximation, which we should not take too seriously.\footnote{Moreover, deriving the \emph{exact} laws of thermodynamics from SM is not a requirement of successful reduction either (see Dizadji-Bahmani et~al. 2010).} Rather than aiming for strict irreversibility, we should expect systems in SM to exhibit what Lavis (2005, 255) calls \textit{thermodynamic-like behaviour} (TD-like behaviour): the entropy of a system that is initially prepared in a low-entropy state increases until it comes close to its maximum value and then stays there and only exhibits frequent small and rare large (downward) fluctuations (contra irreversibility). Even in periods of net entropy increase (such as the moments after the removal of the dividing wall) there can be downward fluctuations (contra monotonicity).

There is a temptation to add to this definition that the approach to equilibrium be fairly quick since some of the most common processes (like the spreading of some gases) are fast. This temptation should be resisted. For one, thermodynamics itself is silent about the speed at which processes take place; in fact, there is not even a parameter for time in the theory! For another, not all approaches to equilibrium are fast. Hot iron cools down slowly, and some systems -- for instance, the so-called Fermi-Past-Ulam system for low energy values -- even approach equilibrium very slowly (Bennetin et~al. 2009). So the approach to equilibrium being fast is not part of a mechanical foundation of thermodynamics. However, it is true, of course, that for SM to be empirically adequate, it has to get relaxation times right. We return to this issue in Section \ref{Further Problems}, where we argue that there is evidence that the relevant systems show the correct relaxation times.

The second respect in which we should require less is universality. The second law of thermodynamics is universal in that it does not allow for exceptions. We should not require the same universality for thermodynamic-like behaviour in SM.
For one, no statistical theory can possibly justify a claim without exceptions; the best one can hope for is to show that something happens with probability equal to one, but zero probability is not impossibility. For another, the relevant systems are time-reversal invariant, and so there will always be solutions that lead from high to low entropy states.\footnote{Conditionalising on the past state \`{a} la Albert (2000) will not make this problem go away because there is no way to rule out that the past state contains solutions that exhibit non-thermodynamic behaviour.} So what we have to aim for is showing that the desired behaviour is \textit{very likely} (Callender 1999). Let $p_{_{TD}}$ be the probability that a system in macrostate $M_{p}$ behaves TD-like. Then what we have to justify is that $p_{_{TD}}\geq 1-\varepsilon$, where $\varepsilon$ is a very small positive real number or zero.

In sum, what needs to be shown is that systems in SM are very likely to exhibit TD-like behaviour. At this point it is important to emphasise that ousting universal and strict irreversibility as the relevant explanandum and replacing it with very likely TD-like behaviour is by no means a trivialisation of the issue. Explaining why systems are likely to behave TD-like is a formidable problem, and the aim of this paper is to propose a solution to it.

Before turning to our positive proposal, let us reflect on the ingredients of such an explanation. In recent years several proposals have been put forward, which aim to justify (something akin to) TD-like behaviour in terms of typicality (see, for instance, Goldstein~2001). TD-like behaviour is said to be typical in dynamical systems, and this fact alone is taken to provide the sought-after explanation.\footnote{For further references and a detailed discussion of this approach, see Frigg (2009b, 2010).} Proponents of this approach reject a justification of TD-like behaviour in terms of ergodicity (to which we turn in the next section), and the context of the discussion makes it clear that they in fact reject (or dismiss as futile) any explanation that makes reference to a dynamical condition (be it ergodicity or something else).

This programme is on the wrong track. It is one of the fundamental posits of Boltzmannian SM that macrostates supervene on microstates. TD-like behaviour is a pattern in the behavior of macrostates; some sequences of macrostates count as being TD-like while others do not. By supervenience, macrostates cannot change without being accompanied by a change in the microstate of the system. In fact, how a macrostate of a system changes is determined by how its microstate changes: the sequence of macrostates of the system is determined by the sequence of microstates. The sequence of microstates depends on the system's initial micro-condition $x$ and the phase flow $\phi_{t}$, which determines how $x$ evolves over the course of time. Hence the dynamics of the macrostates of a system is determined by $\phi_{t}$ and $x$. A fortiori, the phase flow $\phi_{t}$ of the system must be such that it leads to the desired pattern. The central question in the foundations of non-equilibrium SM therefore is: what kind of $\phi_{t}$ give raise to the desired sequence of macrostates? Not all phase flows lead to TD-like behaviour (for instance, a system of harmonic oscillators does not). So the phase flows that lead to TD-like behaviour are a non-trivial subclass of all phase flows on a given phase space, and the question is how this class can be characterised. This question must be answered in a non-question-begging way. Just saying that the relevant phase flows possess a dynamical property called TD-likeness has no explanatory power -- it is a pseudo-explanation of the \emph{vis dormitiva} variety. What we need is a non-trivial specification of a property that only those flows that give raise to TD-like behaviour possess.

It has become customary to discuss the properties of phase flows in terms of Hamiltonians. Phase flows are the solutions to Hamilton's equations of motion, and what sort of motion these equations give raise to depends on what Hamiltonian one inserts into the general equations. So our central question can reformulated as follows: what properties does the Hamiltonian have to posses for the system to behave TD-like?

\section{Ergodic Programmes -- Old and New}\label{Programme}

Boltzmann's original answer to this question was that the relevant Hamiltonians have to be ergodic. This answer has been subjected to serious criticism and has subsequently (by and large) been given up. In this section we introduce the ergodic approach, review the criticisms marshaled against it and outline why these criticisms are either besides the point or can be avoided by appealing to epsilon-ergodicity rather than ergodicity \emph{tout court}.

Consider the phase flow $\phi_{t}(x)$ on $\Gamma_{E}$. The \textit{time-average} of a solution starting at $x\in \Gamma_{E}$ relative to a measurable set $A$ is:

\begin{equation}
L_{A}(x)=\lim_{t\rightarrow\infty}\frac{1}{t}\int_{0}^{t}\chi_{A}(\phi_{\tau}(x))d\tau,
\end{equation}

\noindent where the measure on the time axis is the Lebesgue measure and $\chi_{A}(x)$ is the characteristic function of $A$.\footnote{That is, $\chi_{A}(x)=1$ for $x\in A$ and $0$ otherwise.} Birkhoff's pointwise ergodic theorem ensures that $L_{A}(x)$ exists for all $x$ except, perhaps, for a set of measure zero; i.e., except, perhaps, for a set $B \subseteq \Gamma_{E}$ with $\mu_{E}(B)=0$ (Ott 2002).

Intuitively speaking, a dynamical system is ergodic if and only if (iff) the proportion of time an arbitrary solution stays in $A$ equals the measure of $A$. Formally, $(\Gamma_{E},\mu_{E},\phi_{t})$ is \textit{ergodic} iff for all measurable $A$:
\begin{equation}\label{ergodic}
L_{A}(x)=\mu_{E}(A)\end{equation} for all initial conditions $x\in\Gamma_{E}$ except, perhaps, for in $B$ (which is of measure zero). Derivatively, we say that a \emph{solution} (as opposed to a system) is ergodic iff the proportion of time it spends in $A$ equals the measure of $A$.

If a system is ergodic, it behaves TD-like with  $p_{_{TD}}=1$. Consider an initial condition $x$ that lies on an ergodic solution. The dynamics will carry $x$ to $\Gamma_{M_{eq}}$ and will keep it there most of the time. The system will move out of the equilibrium region every now and then and visit non-equilibrium states. Yet since these are small compared to $\Gamma_{M_{eq}}$, it will only spend a small fraction of time there. Hence the entropy is close to its maximum most of the time and fluctuates away from it only occasionally. Therefore, ergodic solutions behave TD-like. More specifically, as we have seen above, $\mu_{E}$ is a probability measure on $\Gamma_{E}$. This allows us to introduce a probability measure on $\Gamma_{M_{p}}$, $\mu_{p}(C):=\mu_{E}(C)/\mu_{E}(\Gamma_{M_{p}})$ for all $C\subseteq\Gamma_{M_{p}}$, which is the probability that an arbitrary chosen initial condition $x$ lies in set $C\subseteq\Gamma_{M_{p}}$.\footnote{For discussions of interpretations of these probabilities, see Frigg (2009a), Frigg and Hoefer (2010), Lavis (2011) and Werndl (2009c).} The set of `bad' initial conditions (i.e., the ones that are not on ergodic solutions relative to $\Gamma_{M_{eq}}$) in the past state is $B_{p}:=B\cap\Gamma_{M_{p}}$, and from ergodicity it follows that $\mu_{p}(\Gamma_{M_{p}}\setminus B_{p})=1$. We have $p_{_{TD}}=\mu_{p}(\Gamma_{M_{p}}\setminus B)$, and find $p_{_{TD}} = 1$.\footnote{The association of the probability for an initial condition with the Lebesgue measure restricted to $\Gamma_{M_{p}}$ is widely accepted in the current literature; see, e.g., Albert (2000).}

The two main arguments leveled against the ergodic programme are the measure zero problem and the irrelevancy charge. The measure zero problem is that $L_{A}(x)=\mu_{E}(A)$ holds only `almost everywhere', i.e., except, perhaps, for initial conditions of a set of measure zero. This is perceived to be a problem because sets of measure zero can be rather `big' (for instance, the rational numbers have measure zero within the real numbers) and because sets of measure zero need not be negligible if compared with respect to properties other than their measures such as Baire categories (see, e.g., Sklar 1993, 182--88).

This criticism is driven by the demand to justify a strict version of the second law, but this is, as argued in the last section, an impossible goal. The best one can expect is an argument that TD-like behaviour is very likely, and the fact that those initial conditions that lie on non-TD-like solutions have measure zero does not undermine \emph{that} goal. Consequently, we deny that the measure zero problem poses a threat to an explanation of TD-like behaviour in terms of ergodicity. In fact, the solution we propose below is even more permissive in that it allows for sets of `bad' initial conditions that have finite (yet very small) measure.

The second objection, the irrelevancy challenge, is that ergodicity is irrelevant to SM because real systems are not ergodic. This is a serious objection, and the aim of this paper is to develop a response to it. Our solution departs from the observation that less than full-fledged ergodicity is sufficient to explain why systems behave TD-like most of the time. The relevant notion of being `almost but not entirely ergodic' is epsilon-ergodicity.

Intuitively, a dynamical system is epsilon-ergodic iff it is ergodic on the vast majority of $\Gamma_{E}$, namely on a set of measure $\geq 1-\varepsilon$, where $\varepsilon$ is very small real number or zero.\footnote{The concept of epsilon-ergodicity has been introduced into the foundations of SM by Vranas (1998); we comment on his use of it and on how it differs from ours at the end of Section \ref{Evidence}.} To introduce epsilon-ergodicity, we first define the different notion of $\varepsilon$-ergodicity.  $(\Gamma_{E},\mu_{E},\phi_{t})$ is \textit{$\varepsilon$-ergodic}, $\varepsilon\in\field{R},\,0\leq\varepsilon<1$, iff there is a set $Z\subset\Gamma_{E}$, $\mu(Z)=\varepsilon$, with $\phi_{t}(\hat{\Gamma}_{E})\subseteq\hat{\Gamma}_{E}$ for all $t\in\field{R}$, where $\hat{\Gamma}_{E}:=\Gamma_{E}\setminus Z$, such that the system $(\hat{\Gamma}_{E},\mu_{\hat{\Gamma}_{E}},\phi_{t}^{\hat{\Gamma}_{E}})$ is ergodic, where $\mu_{\hat{\Gamma}_{E}}(\cdot):=\mu_{E}(\cdot)/\mu_{E}(\hat{\Gamma}_{E})$ for any measurable set in $\hat{\Gamma}_{E}$ and $\phi_{t}^{\hat{\Gamma}_{E}}$ is $\phi_{t}$ restricted to $\hat{\Gamma}_{E}$. Clearly, a $0$-ergodic system is simply an ergodic system. A dynamical system $(\Gamma_{E},\mu_{E},\phi_{t})$ is \textit{epsilon-ergodic} iff there exists a very small $\varepsilon$ (i.e., $\varepsilon << 1$) for which the system is $\varepsilon$-ergodic.

An epsilon-ergodic system $(\Gamma_{E},\mu_{E},\phi_{t})$ behaves TD-like with $p_{_{TD}}=1-\varepsilon$. Such a system is ergodic on $\Gamma_{E}\setminus Z$, and, therefore, it shows thermodynamic-like behaviour for the initial conditions in $\Gamma_{E}\setminus Z$. If $\varepsilon$ is very small compared to $\mu_{E}(\Gamma_{M_{p}})$, then, by the same moves as explained above for ergodicity, we find $p_{TD}\geq 1-\varepsilon$.\footnote{Notice that the desired result still follows from the weaker premise that $\mu_{E}(\Gamma_{M_{p}}\setminus Z)/\mu_{E}(\Gamma_{M_{p}})$ is close to one.} Hence an epsilon-ergodic system is overwhelmingly likely to behave TD-like.

Our claim is that this explains why real gases behave TD-like because, first, the common no-go results are mistaken (Sections \ref{KAM} and \ref{MM}), and, second, there is good mathematical as well as numerical evidence that systems in SM are in fact epsilon-ergodic (Section \ref{Evidence}).

Before we proceed, we would like mention that ergodicity or epsilon-ergodicity has no implications for how quickly a system approaches equilibrium; i.e., it has not implication for relaxation times. Epsilon-ergodic or ergodic systems can approach equilibrium fast or slow, and which one is the case depends on the particulars of the system. We say more about the relaxation times of the relevant systems in Section \ref{Further Problems}.

\section{The KAM Results and Increasing Perturbation Parameters}\label{KAM}

\subsection{The Kolmogorov-Arnold-Moser theorem}\label{The Kolmogorov-Arnold-Moser theorem}

Support for the irrelevancy charge is mustered by appeal to two theorems, the KAM-theorem and the Markus-Meyer theorem, which are taken to show that systems in SM are not ergodic. We discuss the former in this section and latter in Section~6. Our main contention is that the arguments marshaled against ergodicity based on the KAM-theorem rest on a misinterpretation: the KAM-theorem is irrelevant since gases in SM do not satisfy the premises of the theorem.

Let us now discuss the KAM-theorem. A function $F$ is a \textit{first integral} of a dynamical system iff its Poisson bracket $\{H,F\}$ is  equal to zero, where $H$ is the Hamiltonian of the system; in physical terms a first integral is a constant of motion. A dynamical system with $n$ degrees of freedom is \textit{integrable} (in the sense of Liouville) iff the system has $n$ independent first integrals $F_{i}$ and these integrals are in involution (i.e., $\{F_{i},F_{j}\}=0$ for all $i,j,\,1\leq i,j\leq n$). A dynamical system is \textit{nonintegrable} iff it is not integrable. For an integrable system the energy hypersurface is foliated into tori, and there is periodic or quasi-periodic motion with a specific frequency on each torus (cf., Arnold~1980;  Arnold~et~al.~1985).

The Kolmogorov-Arnold-Moser theorem (KAM-theorem)  describes what happens when an integrable Hamiltonian system is perturbed by a very small nonintegrable perturbation, i.e., what happens with the Hamiltonian $H=H_{0}+\lambda H_{1}$, where $H_{0}$ is integrable, $H_{1}$ is nonintegrable and $\lambda>0$ is a very small perturbation parameter. The theorem says that under certain conditions\footnote{Intuitively, these conditions say that for $H_{0}$ the ratios of frequencies on a given torus vary smoothly from torus to torus. Technically, \label{iso} the conditions are that (i) one of the frequencies never vanishes, and (ii) that the ratios of the remaining $n-1$ frequencies to the non-vanishing frequency are always functionally independent on the energy hypersurface.} the following holds on the hypersurface of constant energy: the tori with sufficiently irrational winding numbers (i.e., frequency ratios) survive the perturbation; this means that the solutions on these tori behave like the ones in the integrable system and are thus stable. The other tori, which lie between the stable ones, are destroyed, and here the motion is irregular. Furthermore, the measure of the regions which survive the perturbation goes to one as the perturbation parameter goes to zero.

The region on $\Gamma_{E}$ in which the tori survive and the region in which they break up are both invariant under the dynamics. The motion on the region with surviving tori cannot be ergodic (or epsilon-ergodic) because the solutions are confined to tori. Therefore, dynamical systems to which the KAM-theorem applies are not ergodic, and for a small enough perturbation, they are not epsilon-ergodic either (Arnold 1963; Arnold~et~al.~1985).

This implication of the KAM-theorem is often taken to warrant the conclusion that many (if not all) systems in SM fail to be ergodic:
\begin{quote}
[T]he evidence against the applicability [of ergodicity in SM] is strong. The KAM-Theorem leads one to expect that for systems where the interactions among the molecules are non-singular, the phase space will contain islands of stability where the flow is non-ergodic. (Earman and Redei 1996)
\end{quote}
\begin{quote}
Actually, demonstrating that the conditions sufficient for the regions of KAM-stability to exist can only be done for simple cases. But there is strong reason to suspect that the case of a gas of molecules interacting by typical intermolecular potential forces will meet the conditions for the KAM result to hold. [...] So there is plausible theoretical reason to believe that more realistic models of typical systems discussed in statistical mechanics will fail to be ergodic. (Sklar 1993)
\end{quote}

First appearances notwithstanding, the KAM-theorem does not establish that relevant systems in SM are not ergodic (and \emph{a fortiori} it does not establish that they are not epsilon-ergodic). To see why, attention has to be paid to the fine print of the theorem. The important -- and often ignored -- point is that the KAM-theorem only applies to \emph{extremely small} perturbations. Percival makes this point vividly for systems with two degrees of freedom:

\begin{quote}
Arnold's proof only applies if the perturbation is less than $10^{-333}$ and Moser's if it is less than $10^{-48}$, in appropriate units. The latter is less than the gravitation perturbation of a football in Spain by the motion of a bacterium in Australia! The KAM proofs were a vital contribution to dynamics because they showed that regular motion is not effectively restricted to integrable systems, but numerically they are not yet of practical value. (Percival 1986)
\end{quote}

So the applicability of the KAM-theorem to realistic two particles systems is severely limited. Most important for our purposes is that in SM the applicability of the KAM-theorem fails drastically. For many classes of systems in SM the largest admissible perturbation parameters have been calculated, and one finds that they rapidly converge toward zero as $n$ tends toward infinity (Pettini 2007). Hence, as Pettini points out, ``for large $n$-systems -- which are dealt with in statistical mechanics -- the admissible perturbation amplitudes for the KAM-theorem to apply drop down to exceedingly tiny values of no physical meaning'' (Pettini and Cerruti-Sola 1991; Pettini 2007, 60). If the perturbation is larger, the surviving tori disappear and, at least in principle, nothing stops the motion from being epsilon-ergodic (or even ergodic).
Moreover, gases in SM are not even moderately small perturbations of integrable systems. An ideal gas (a collection of non-interacting particles) is integrable, but even a dilute real gas cannot be represented as a small perturbation of an ideal gas. As we will see in Section \ref{Evidence}, particles in real gases repel each other strongly when they come close to each other. Hence the perturbation parameter is comparatively large (Penrose and Lebowitz 1973). To sum up, the KAM-theorem cannot be expected to apply to realistic systems in SM. Even the smallest interactions, e.g., such as interactions between molecules, introduce perturbations far greater than the one allowed by the KAM-theorem, which is therefore simply silent about what happens in such systems.

Furthermore, the class of systems which can be represented as a  perturbation of an integrable system is very special (the KAM-theorem deals with a subclass of this class -- those with extremely small perturbation parameters). And it is at best unclear whether many systems in SM fall within that class.\footnote{Thanks to Pierre Lochak for pointing this out to us.}
We conclude that the KAM-theorem does not show that systems in SM fail to be ergodic (or epsilon-ergodic), and one cannot dismiss the ergodic approach by appealing to the KAM-theorem.

\subsection{Arnold Diffusion and Increasing Perturbation Parameters}\label{Arnold Diffusion}

As argued, systems in SM cannot be represented as very small perturbations of integrable systems; yet it may be that at least some systems in SM are a moderate or larger nonintegrable perturbation of an integrable system. So we have to understand how such systems behave. This is best achieved by studying what happens if the perturbation parameter of a system to which the KAM theorem applies is increased. We will see that there is evidence that the motion is epsilon-ergodic.

Let us begin by considering the motion for very small perturbations (i.e., the KAM-regime) because this will lead to a better understanding of what happens when the perturbation parameter is increased. For very small perturbation parameters, most of the energy hypersurface is taken up by regular motion. The region of irregular motion is of very small measure, but, interestingly, it is nevertheless always everywhere dense on the energy hypersurface. Now in systems with two degrees of freedom, invariant tori separate different irregular regions from each other because solutions remain `trapped' between two tori (this is usually illustrated in the two-dimensional Poincar\'{e} section of a system). The irregular motion in such a system cannot possibly form a single connected region; thus the flow cannot diffuse or be ergodic on that region. However, the situation completely changes for systems with three or more degrees of freedom (the cases relevant to SM). The  energy  hypersurface is $2n-1$ dimensional, and another surface must be of $2n-2$ dimensions to divide it into two disconnected parts.  But since the invariant tori are of dimension $n$ and $2n-2>n$ for all $n>2$, the invariant tori do not divide the energy hypersurface into separate parts for $n>2$: the stable tori are like circles in a three-dimensional Euclidean space. Hence the irregular motion (which, recall, is everywhere dense on the energy hypersurface) can form a singly connected region, commonly referred to as a \emph{web} (or \emph{Arnold web}).

Arnold (1963, 1994) conjectured that for any extremely small $\lambda$ and generic Hamiltonian perturbations $H_{1}$ the following holds on the hypersurface of constant energy: for any two tori $T$ and $T'$ there is a solution connecting an arbitrary small neighbourhood of $T$ with an arbitrary small neighbourhood of $T'$. Intuitively speaking, this conjecture says that there is diffusion on the energy hypersuface along all the different tori.\footnote{That is, there is diffusion relative to the action variables describing the torus on which a state is located.} This diffusion for extremely small perturbation parameters is called \textit{Arnold diffusion}.

Arnold's hypothesis has not been proven in full generality, but  there are good reasons to believe that it is true. First, Arnold diffusion has been proven to exist in many concrete examples (see, e.g., Arnold 1964;  Berti et al.~2003; Delshams and Huguet 2009; Mather~2004). Furthermore, numerical studies confirm the existence of an Arnold web for arbitrary small perturbations of integrable Hamiltonian systems. For instance, Froeschl\'{e} et~al. (2000) and Fr\"{o}schle and Schneidecker (1975) have shown that for very small perturbation parameters there appears to be a single web of unstable motion. Moreover, the motion restricted to the irregular web appears to be ergodic Ott (2002, 257).

The results about Arnold diffusion are crucial because they  are the only strict mathematical results showing that there is diffusion on the irregular region of perturbed integrable systems.
For our concern these results are important because when the perturbation parameter $\lambda$ is increased, it is found that exactly this irregular region grows larger and larger. Although there exist no mathematically rigorous results any longer in that regime, it is widely believed that there is diffusion on this irregular region similar to the one for very small perturbations. And the fact that diffusion is strictly proven for very small perturbations makes it plausible that there is also diffusion for larger perturbations.\footnote{For our concerns it is also noteworthy that Arnold diffusion shows that even extremely \textit{regular} systems, namely arbitrary small perturbations of integrable systems, show random motion in the sense that there is an everywhere dense web on which there is diffusion.}
 This is backed up by numerical investigations, which suggest that, as the parameter gets larger, more and more of the invariant tori break up, and the region with irregular motion covers larger and larger parts of the energy hypersurface. For perturbations higher than a specific moderate perturbation, nearly all or all of the energy hypersurface seems to be taken up by irregular motion, and hence the motion appears to be epsilon-ergodic (Chirikov 1979, 1991; Froeschl\'{e} et al.~2000; Ott 2002; Vivaldi 1984).
It could be the case that very small islands of regular motion persist even for arbitrary large perturbations. But then these regular regions are very small, and so while the system would fail to be ergodic, it will still be epsilon-ergodic. Furthermore, there is evidence that, everything else being equal, the main region of ergodic behaviour grows larger and larger as the number of degrees of freedom increases (Fr\"{o}schle and Schneidecker~1975; Reidl and Miller 1993; Sz\'{a}sz 1996).

In sum, for moderate or larger perturbations of integrable systems the motion appears to be epsilon-ergodic on the entire energy hypersurface. It might be that at least some systems in SM can be represented as a moderate or larger nonintegrable perturbation of integrable systems; and then these systems can be expected to be epsilon-ergodic, which is what we need.

\section{The Markus-Meyer Theorem}\label{MM}

We now turn to the second main argument against ergodicity, which is based on the Markus-Meyer theorem (MM-theorem) (Markus and Meyer 1974).

First of all, we need to introduce the central concepts of a topology on a function space and a generic Hamiltonian. Consider a class $\Lambda$ of functions of a certain kind on a set $M$. In order to be able to say that some functions are closer to each other than others we introduce a \textit{topology} on $\Lambda$. If $\Lambda$ consists of infinitely differentiable Hamiltonians on a set $M$, it is common to choose the so-called Whitney topology, according to which two Hamiltonians are close just in case the absolute value between their vector fields and all their derivatives is small.\footnote{Intuitively the Whitney topology can be characterised as follows. Consider two infinitely differentiable Hamiltonians $H_{1}$ and $H_{2}$ on $M$, $H_{1}^{'}$, $H_{2}^{'}$, $H_{1}^{''},\,\,H_{2}^{''}$, etc. being their derivatives. $H_{1}$ and $H_{2}$ are close just in case $|H_{1}-H_{2}|$, $|H_{1}^{'}-H_{2}^{'}|$, etc.\ are all small (cf.\ Hirsch 1976).}

We can now define the notion of a generic function. A set is \textit{meagre} iff it is the countable union of nowhere dense sets; and a set is \textit{comeagre} iff its complement is meagre (cf. Oxtoby 1980). Loosely speaking, a meagre set is the topological counterpart of the measure-theoretic notion of a set of measure zero, and a comeagre set is the counterpart of a set of measure one.\footnote{There is also a classification of sets into first and second Baire category (see~Sklar~1993, 182--88). A meagre set is the same as a set of first Baire category. Sets of second Baire category are defined as sets which are not of first Baire category. This means that being of second category is different from being comeagre. There are sets which are neither meagre nor comeagre (they are the topological counterpart of sets between measure zero and one), but, by definition, they are of second Baire category.} So generic functions of $\Lambda$ are ones that belong to $\bar{\Lambda}\subseteq\Lambda$ where $\bar{\Lambda}$ is comeagre.\footnote{Function spaces cannot be equipped with measures that one could use to define the notion of generic functions measure-theoretically. Thus this notion is always defined topologically.}

Consider the function space $\Lambda$ of all infinitely differentiable Hamiltonians on a compact space $M$; the topology is induced by the Whitney topology on all potential functions.\footnote{What follows also holds relative to the Whitney topology of the Hamiltonians. However, it is physically more natural to vary only the potential functions and not the expression for the kinetic energy (Markus and Meyer 1969, 1974).} An (epsilon-) ergodic  \emph{Hamiltonian} (as opposed to a flow or a solution) is one which has a dense set of energy values for which the flow on the hypersurface of constant energy is (epsilon-) ergodic.
The MM-theorem says that the set of Hamiltonians in $\Lambda$ which are \emph{not} ergodic is comeagre; in other words, non-ergodic Hamiltonians are generic. Furthermore, a closer look at the proof of the theorem shows that it implies that the set of Hamiltonians in $\Lambda$ which fail to be epsilon-ergodic is also comeagre and hence generic (Markus and Meyer 1969, 1974; see also Arnold~et~al.~1985, 193).

And things seem to get worse. It is a plausible demand that physical properties be robust under small structural perturbations. In our case this amounts to requiring that if a system is (epsilon-) ergodic, a system with a very similar potential function should be epsilon-ergodic as well. In technical terms, one would expect that for any (epsilon-) ergodic Hamiltonian $H$ there is an open set in $\Lambda$ around $H$ such that all Hamiltonians in the open set are (epsilon-) ergodic as well. The MM-theorem says that (epsilon-) ergodic system are not stable in that sense because non-epsilon-ergodic systems are dense in $\Lambda$. So the MM-theorem seems to imply that (epsilon-) ergodic systems are not structurally stable, which rules out (epsilon-) ergodicity as property with explanatory power. For this reason Emch and Liu (2002) call the Markus-Meyer theorem ``the main no-go theorem'' for the ergodic approach.

On the face of it these arguments seem devastating, both to the original ergodic programme and to our rendition based on epsilon-ergodicity. We now argue that this impression unravels under a closer analysis.

Take first the proposition that generic Hamiltonians are not (epsilon-) ergodic, and that therefore (epsilon-) ergodicity is immaterial to the foundation of SM.\footnote{There is also a question whether being generic is a reasonable requirement to begin with. What matters is whether those systems \emph{actually} studied in SM are epsilon-ergodic, and whether or not these are generic seems to be immaterial.} This argument fails on two counts. First, the MM-theorem only applies to compact phase spaces; compactness is essential in the proof and the proof does not go through  for non-compact spaces. But nearly all systems considered in classical mechanics have non-compact phase spaces (see, e.g., Arnold 1980). One might be tempted to reply that this can easily be resolved: since it is the flow on the energy hypersurface that we are ultimately interested in, and since the energy hypersurface typically \emph{is} compact, we simply apply the theorem to the energy hypersurface. This cannot be done. The theorem is about the \emph{full} phase space $\Gamma$ of a system and cannot be rephrased as a theorem about energy hypersurfaces. This is because the theorem treats the energy of the system as a free parameter, and the essential result is derived by varying the value of that parameter. This simply makes no sense on an energy hypersurface where, by definition, the energy is held constant.

Second, even if the theorem were applicable, once we understand \emph{how} the main proposition of the theorem is established, it becomes clear that it fails to establish the irrelevancy of (epsilon-) ergodicity for gases in SM. As we have seen above, the definition of an (epsilon-) ergodic Hamiltonian used in the MM-theorem is that there is a dense set of energy values for which the motion on the energy hypersurface is (epsilon-) ergodic. This implies that (epsilon-) ergodicity is required \textit{arbitrarily close to any possible energy value}. Hence a Hamiltonian is non-ergodic if there exists only \emph{one} value for which this is not the case. Proving that there is one such value is the strategy of the theorem: Markus and Meyer prove that for generic Hamiltonians there is exactly one minimal value of the energy (where the motion is a general elliptic equilibrium point), and then show  that for energy values which are arbitrarily close to this minimum the motion on the energy hypersurface is not epsilon-ergodic.

However, it is doubtful that these very low energy values are relevant to the behaviour of gases. For many systems in SM for energy values close to the minimum value of the energy the classical-mechanical description breaks down because quantum-mechanical effects come in. Then it is irrelevant if the systems are not epsilon-ergodic for these energy values (Penrose 1979; Reichl~1998, 488, Vranas~1998). Even when they are physically relevant, the values close to a minimum value of the energy are irrelevant to the behaviour of \textit{gases}, and thus the theorem has no damaging effect: these low energy values, to the best of our knowledge, never correspond to gases, but to glasses or solids.\footnote{For instance, numerical evidence suggests that for several systems there is a liquid-glass transition which goes hand in hand with a transition from epsilon-ergodic to non-epsilon-ergodic behaviour (De~Souza and Wales, 2005).} And for larger energies, numerical evidence suggests that the motion is indeed epsilon-ergodic. This point is also emphasised by Ford and Stoddard (1073, 1504) (but not with reference to the Markus-Meyer theorem, which had not been published then): ``nothing precludes the existence of a cricial energy, depending perhaps on various system parameters, above which systems with attractive forces are no less ergodic than the hard-sphere gas''.\footnote{The sceptic might now try to prove a theorem analogous to the Markus-Meyer theorem with a weaker notion of an epsilon-ergodic Hamiltonian, requiring only that there is epsilon-ergodicity for a somewhere dense set of energy levels. However, Markus and Meyer (1969, 1974) show that such a theorem is false.}

It is a corollary of this analysis that the stability challenge has no bite either. As we have just seen, Hamiltonians fail to be ergodic because things go wrong close to the minimum energy, while the theorem is silent about higher energies. So if we consider an ergodic gas at a realistic energy, it does not follow from the theorem that disturbing it a little bit would result in a non-ergodic system. The theorem is silent about what happens in such a situation.

In sum, the MM-theorem poses no threat to an explanation of TD-like behaviour of gases in terms of (epsilon-) ergodicity.

\section{Relevant Cases}\label{Evidence}

So far we have shown that arguments against ergodicity in SM unravel under careful analysis. Yet in order to render an explanation of TD-like behaviour based on epsilon-ergodicity plausible, more is needed than showing that no-go theorems have no force. We need positive arguments for the conclusion that gases are indeed epsilon-ergodic. The aim of this section is to provide such arguments.

Given the intricacies of the mathematics of dynamical systems, it will not come as a surprise that rigorous results are few and far between. In the absence of far-reaching general results our strategy is one of piecemeal and focused analysis. Our analysis is focused because we consider only physically relevant systems; it is piecemeal because we study the systems individually and look both at the available mathematical and numerical evidence. The conclusion is that there are good reasons to believe that relevant systems are epsilon-ergodic.

The dynamics of a gas is specified by the potential which describes the inter-particle forces. Two potentials stand out: \textit{the Lennard-Jones potential} and \textit{the hard-sphere potential.} The hard-sphere potential models molecules as impenetrable spheres of radius $R$ that bounce off elastically. It is the simplest potential, which is why it is widely used in both mathematical and numerical studies. For two particles it has the form
\begin{equation}
U(r)=\infty\,\,\,\textnormal{for}\,\,\,r<R\,\,\,\textnormal{and 0 otherwise},
\end{equation}
where $r$ is the distance of the particles. The potential of the entire system is obtained by summing over all two-particles interactions. The hard-sphere potential simulates the steep repulsive part of realistic potentials (McQuarrie 2000, 234).

The Lennard-Jones potential for two particles is
\begin{equation}
U(r)=4\alpha\left(\left(\frac{\rho}{r}\right)^{12}-\left(\frac{\rho}{r}\right)^{6}\right),
\end{equation}
where $r$ is the distance between two particles, $\alpha$ is the depth of the potential well and $\rho$ is the distance at which the inter-particle potential is $0$. This function is in good empirical agreement with data about inter-particle forces (McQuarrie 2000, 236--37; Reichl 1998, 502--05). The potential of the entire system is then obtained by summing over all two-particle-interactions or by considering only the nearest neighbour interactions.

Let us begin with a discussion of the hard-sphere potential. It was already studied by Boltzmann (1971), who conjectured that hard-sphere systems show ergodic behaviour when the number of balls is large. From a mathematical viewpoint it is easier to deal with particles moving on a \textit{torus}, rather than particles moving in a box with hard walls (or in other containers with hard walls). For a hard-sphere moving on a torus there are no walls; it is as if a ball, when reaching the wall of the box, reappears at the opposite side instead of bouncing off.\footnote{Because there are no walls, the motion on a torus has a second constant of motion next to the energy: total momentum. Hence, in this case, the question of interest is whether the motion is ergodic relative to a given value of the energy and a given value of the total momentum. The results referred to in this section add up to an affirmative answer for almost all parameter values. As soon as there are walls, total momentum ceases to be an invariant. That the motion of hard balls on a torus is ergodic is an important result: it gives us reason to expect that the
motion of hard balls will also be ergodic when there are walls.} Studying the motion of hard-spheres on a torus is not an irrelevant mathematical passtime. If anything, walls render the motion more and not less random, and hence establishing that the motion on torus is ergodic supports the conclusion that the motion in box is ergodic too.\footnote{For a discussion of this point in the case of the stadium billiards see Chernov and Markarian (2006).}
Sinai (1963) conjectured that the motion of $n$ hard-spheres on $T^{2}$ and on $T^{3}$ is ergodic for all $n\geq 2$ where $T^{m}$ is the $m$-torus(cf.\ Sz\'{a}sz 1996), a hypothesis now known as the `Boltzmann-Sinai ergodic hypothesis'. Sinai (1970) made the first significant step towards a rigorous proof of this hypothesis when he showed that the motion of $2$ hard-spheres on $T^{2}$ is ergodic.\footnote{All the cases of hard-sphere systems which are reported in this section to be ergodic are even  Bernoulli systems (i.e., they are strongly chaotic). For a discussion of Bernoulli systems, see Werndl~(2009a, 2009b, 2011).} Since then a series of important results have been obtained, which, taken together, add up to an almost complete proof of the Boltzmann-Sinai ergodic hypothesis (and mathematicians who work in this field and know about the progress in the last years think that a full proof can be expected to be forthcoming soon -- see Simanyi 2009). The following three results are particularly noteworthy. First, Simanyi (1992) proved that the motion of $n$ hard-spheres on $T^{m}$ is ergodic for all $m\geq n$, $n\geq 2$. Second, Simanyi (2003) proved that systems of $n$ hard spheres are ergodic on $T^{m}$ for all $n\geq 2$, all $m\geq 2$ and for almost all values $(m_{1},\ldots,m_{n},r)$, where $m_{i}$ is the mass of the $i$-th ball and $r$ is the radius of the balls.\footnote{Unfortunately, no effective method is known of checking whether a given $(m_{1},\ldots,m_{n},r)$ is among this set of almost all values (implying that the system is proven to be ergodic). This means that we do not know whether the system is ergodic for equal masses (the case we are mainly interested in) (Simanyi 2009, 383).}
Third, Simanyi (2009) proved that systems of $n$ hard spheres are ergodic on $T^{m}$ for all $n$ and all $m$ provided that the Sinai-Chernov Ansatz is true.\footnote{Let $M$ be the set of all possible states of the hard-sphere system, and consider $\partial M$, the boundary of $M$. Let $SR^{+}$ consist of all states $x$ in $\partial M$ corresponding to singular reflections with the post-collision velocity $v_{0}$, for an arbitrary $v_{0}$. The Chernov-Sinai Ansatz postulates that for for almost every $x\in SR^{+}$ the forward solution originating from $x$ is geometrically hyperbolic (Simanyi 2009, 392).} This Ansatz is known to hold for systems that are similar (from a mathematical viewpoint) to systems of three or more hard balls (cf., Simanyi 2003, 2009; Szasz 1996). Furthermore, as just mentioned, hard ball systems are proven to be ergodic for almost all parameter values, and, as we will see in the next paragraph, there is numerical evidence that hard ball systems are ergodic. For this reason it is plausible to assume that the Sinai-Chernov Ansatz is true (an assumption which is generally accepted among mathematicians).

The more realistic case of the motion of hard-spheres in a box (rather than on a torus) is extremely difficult, and fewer analytical results have been obtained. Simanyi (1999) proved that the motion of two balls in an $m$-dimensional box is ergodic for all $m$. There are good reasons to believe that the same result holds for a arbitrary number of balls because the behaviour of hard spheres in a box is at least as random as the behaviour of hard spheres moving on a torus and the latter have been proven to be ergodic for almost all parameter values. This conclusion is supported by numerical simulations. Zheng et al.\ (1996)  investigated the motion of identical hard-spheres in a two-dimensional and a three-dimensional box, and the behaviour was found be ergodic. And Dellago and Posch (1997) found evidence that a large number of identical hard-spheres in a three-dimensional box show ergodic behaviour for a wide range of densities.

Let us now turn to the Lennard-Jones potential, where things are more involved. Donnay (1999) proved that for some values of the energy a system of two particles moving on $T^{2}$ under a generalised Lennard-Jones type potential is not ergodic.\footnote{Generalised Lennard-Jones potentials include both potentials of the same general shape as Lennard-Jones potentials and a much broader class of potentials. More specifically, to make the mathematical treatment easier, Donnay (1999) assumes that a generalised Lennard-Jones potential has only finite range; that is, there is an $R>0$ such that $U(r)=0$ for all $r\geq R$. Apart from this, a generalised Lennard-Jones potential is defined to be a smooth potential where (i) the potential is attracting for large $r$, and (ii) the potential approaches infinity at some point as $r$ goes to zero.}
This result illustrates that ergodicity can be destroyed by replacing inelastic collisions by an everywhere smooth potential. Yet it leaves open what happens for a large number of particles (generally, the larger the number of particles, the more likely a system is to be ergodic). Now it is important to note that even if systems with Lennard-Jones potentials and with a large number of particles turn out to be non-ergodic, they appear to be epsilon-ergodic (Stoddard and Ford 1973), as also expressed by Donnay:
\begin{quote}
Even  if  one  could  find  such  examples [generalised Lennard-Jones systems with a large number of particles that are non-ergodic], the measure of the set of solutions constrained to lie near the elliptic periodic orbits is likely to be very small. Thus from a practical point of view, these systems may appear to be ergodic. (Donnay 1999, 1024)
\end{quote}

This is backed up by numerical studies on Lennard-Jones potentials, which have shown the following. The system has an energy threshold (a specific value of the energy) such that for values above that energy threshold the motion of the system appears to be epsilon-ergodic and for values below that threshold the system appears not to be epsilon-ergodic (Bennetin et al.\ 1980; Bocchieri et al.\ 1970; Diana et al.\ 1976; Stoddard and Ford 1973). What matters here is that the energy values below the energy threshold are very low. As a consequence, quantum effects will play a role and the classical SM description will not adequately describe the physical systems in question (Penrose 1979; Reichl~1998, 488; Vranas 1998). Therefore, the behaviour of the systems with energy values below the threshold are irrelevant to this paper. To conclude, the evidence supports the claim that Lennard-Jones type systems are epsilon-ergodic for the relevant energy values.

Finally, after the discussion of the hard-sphere potential and the Lennard-Jones potential, let us briefly mention some of the most important mathematical and numerical results about other potentials relevant to SM. First, Donnay and Liverani (1991) proved that two particles moving on $T^{2}$ are ergodic for a wide class of potentials, namely for a general class of repelling potentials, a general class of attracting potentials, and a class of mixed potentials (i.e., attracting and repelling parts). Particularly remarkable here is that the mixed potentials are \textit{everywhere smooth}. Everywhere smooth potentials are usually regarded as more realistic than potentials which involve singularities. And Donnay and Liverani (1991) were the first to mathematically prove that some everywhere smooth Hamiltonian systems are ergodic. Second, one of the most extensive numerical investigations of systems with many degrees of freedom relevant to SM has been about a one-dimensional self-gravitating system consisting of $n$ plane-parallel sheets with uniform density (this system is of interest in plasma physics). Strong evidence was found that the measure of the ergodic region increases rapidly with $n$ and that for $n\geq 11$ the system is completely ergodic (Fr\"{o}schle and Schneidecker 1975; Reidl and Miller 1993; Wright and Miller 1984).

In sum, there is good evidence that all gases in SM are epsilon-ergodic, and, crucially, there is no known counter-instance. Before turning to further issues, we would like to compare our own conclusion to Vranas' (1998), who (as we briefly mentioned above) introduced the notion of epsilon-ergodicity into the foundations of SM and argued that relevant systems are indeed epsilon-ergodic. Our strategy shares much in common with his -- in particular the focus on physically relevant cases. Yet there are important differences. The most significant difference is that we consider Boltzmannian non-equilibrium theory while he studies Gibbsian equilibrium theory. There are also differences in the choice of the `inductive base': we have tried to give as fully as possible an account of currently available analytical results, while Vranas focusses almost entirely on numerical studies. Furthermore, the scope of our argument is different. Vranas sees the cases he discusses as supporting the hypothesis that \emph{all} non-integrable Hamiltonian systems with many degrees of freedom are epsilon-ergodic (see, e.g., \emph{ibid.}, 697). We restrict our claim to gases because (as we point out in the next section) there are systems -- most notably solids -- whose dynamics does not seem to be epsilon-ergodic. Such systems demand a different analysis.

\section{Further Issues on the Horizon}\label{Further Problems}

To conclude our discussion, we want to address two potentially problematic points. The first concerns the question of relaxation times; the other is that some systems behave TD-like and yet fail to be ergodic.

Epsilon-ergodicity itself implies nothing about the speed at which systems approach equilibrium. However, to be empirically adequate, SM needs to predict correct relaxation times. For our approach this means that the relevant systems in SM, in addition to being epsilon-ergodic, must show the correct relaxation times. Unfortunately, rigorous results about relaxation times are even harder to come by than ergodicity proofs; in fact, from a strictly mathematical viewpoint nothing is known about the relaxation times of systems in SM (Chernov and Young 2000). However, several numerical studies provide evidence that both hard-sphere and Lennard-Jones gases approach equilibrium at the right speed. First, Dellago and Posch (1997) and Zheng et al.\ (1996, 3249 and 3251) consider the evolution of both the position and momentum variables for hard-sphere gases and find that equilibrium is reached after only a few mean collision times. Second, Bocchieri et~al. (1970) and Yoshimura (1997) show that Lennard-Jones gases approach equilibrium very quickly, namely in less than $10^{-8}$ seconds.\footnote{These studies investigate the relaxation to energy equipartition, indicating the approach to equilibrium.} Moreover, to the best of our knowledge, there are no numerical studies indicating that the relevant gases do not approach equilibrium very quickly.\footnote{Furthermore, for KAM-type systems the diffusion on the regions of irregular behaviour becomes faster as the perturbation parameter increases (Chirikov 1979, 1991; Froeschl\'{e} et al.~2000; Ott 2002; Vivaldi 1984), which also suggests that realistic gases converge fast (cf., Section~\ref{KAM}).} Hence while the issue of relaxation times certainly deserves more attention than it has received so far, currently available results indicate that the relevant systems behave as expected.

The second issue concerns the alleged non-necessity of (epsilon-) ergodicity for TD-like behaviour. Attention is often drawn to particular systems that fail to be ergodic and yet behave TD-like, from which it is concluded that ergodicity cannot explain TD-like behaviour. Common `counter-instances' are the following. First, in a solid the molecules oscillate around fixed positions in a lattice, and as a result the phase point of the system can only access a small part of the energy hypersurface (Uffink 2007, 1017). Yet solids behave TD-like. Second, the Kac Ring Model is not ergodic while exhibiting TD-like behaviour (Bricmont 2001). Third, a system of $n$ uncoupled anharmonic oscillators of identical mass is not ergodic but still behaves TD-like (\emph{ibid}.). Fourth, a system of non-interacting point particles is known not be ergodic; yet this system is often studied in SM (Uffink 1996, 381).

First appearances notwithstanding, these examples do not undermine our claim that epsilon-ergodicity explains TD-like behaviour in gases. The Kac-ring model and uncoupled harmonic oscillators have little if anything to do with gases, and hence are irrelevant. The ideal gas has properties very different from those of real gases because there are no collisions in an ideal gas and collisions are essential to the behaviour of gases. So while the ideal gas may be an expedient in certain context, no conclusion about the dynamics of real gases should be drawn from it.

Needless to say, solids are no gases either and hence do not bear on our claim. However, the case of solids draws our attention to an important point, namely that an explanation of TD-like behaviour in terms of epsilon-ergodicity cannot be universal. Some solids not only fail to be ergodic; they also fail to be nearly ergodic. For this reason one cannot explain thermodynamic-like behaviour for all solids in terms of epsilon-ergodicity. This highlights that further work is needed: explaining thermodynamic-like behaviour for solids is an unsolved problem, and one which deserves more attention than it has received so far; in fact, even the Boltzmannian macrostate structure of solids is unknown!

At present it is not know whether epsilon-ergodicity will play a role in such an explanation, and if it does what that role will be. But we do not see this as a problem for the current project. Epsilon-ergodicity explains TD-like behaviour in gases, and should it turn out not to explain TD-like behaviour in other materials that does not undermine its explanatory power in the relevant domain. Two scenarios seem possible. The first is that epsilon-ergodicity will turn out to be a special case of a (yet unidentified) more general dynamical property that \emph{all} systems that behave TD-like posses. In this case epsilon-ergodicity turns out to be part of a general explanatory scheme. The other scenario is that there is no such property and the best we come up with is a (potentially long) list with different dynamical properties that explain TD-like behaviour in different cases. But nature turning out to be disunified in this way would be no reason to declare explanatory bankruptcy: `local' explanations are explanations nonetheless!

\section{Conclusion}\label{Conclusion}
The aim of this paper was to explain why gases exhibit thermodynamic-like behaviour. The canonical answer, originally proffered by Boltzmann, is that the systems have to be ergodic. In this paper we argued that some of the main arguments against this answer, in particular, arguments based on the KAM-theorem and the Markus-Meyer theorem, are beside the point or inconclusive. Then we argued that something close to Boltzmann's original proposal is true: gases show thermodynamic-like behaviour when they are epsilon-ergodic, i.e., ergodic on the entire accessible phase space except for a small region of measure epsilon; and there are good reasons to believe that the relevant physical systems are epsilon-ergodic. Consequently, for gases epsilon-ergodicity seems to be the sought-after explanation of thermodynamic-like behaviour.

\section*{Acknowledgements}
Earlier version of this paper have been presented at the 2010 BSPS conference, PSA 2010, and at the philosophy of physics research seminar in Oxford; we would like to thank the audience for valuable discussions. We also want to thank Scott Dumas, David Lavis, Pierre Lochak and David Wallace for helpful comments. Roman Frigg also wishes to acknowledge supported from the Spanish Government research project FFI2008-01580/CONSOLIDER INGENIO CSD2009-0056.

\section*{References}

\begin{list}{}{    \setlength{\labelwidth}{0pt}
    \setlength{\labelsep}{0pt}
    \setlength{\leftmargin}{24pt}
    \setlength{\itemindent}{-24pt}
  }

\item Albert, David 2000. {\em Time and Chance}. Cambridge/MA and London: Harvard University Press.

\item Arnold, Vladimir I. 1963. ``Small Denominators and
  Problems of Stability of Motion in Classical and Celestial Mechanics.'' {\em
  Russian Mathematical Surveys} 18:85--193.

\item 
--------- 1964. ``Instabilities in
  Dynamical Systems With Several Degrees of Freedom.'' {\em Soviet Mathematics
  Doklady} 5:581--585.

\item 
--------- 1980. {\em Mathematical
 Methods of Classical Mechanics}. New York, Heidelberg, Berlin: Springer.

\item 
--------- 1994. ``Mathematical Problems
  in Classical Physics.'' In \emph{Trends and Perspectives in
  Applied Mathematics}, ed. Lawrence Sirovich, 1--20. Berlin and New York: Springer.

\item Arnold, Vladimir I., Valery I. Kozlov, and Anatoly I. Neishtat 1985. {\em Dynamical Systems {III}}. Heidelberg: Springer.

\item 
Bennetin, Giancarlo, Roberto Livi, and Antonio Ponno  2009. ``The {Fermi-Pasta-Ulam} Problem: Scaling Laws Vs.
  Initial Conditions', {\em Journal of Statistical Physics} 135:873--893.

\item 
Bennetin, Giancarlo, Guido Lo~Vecchio, and Alexander Tenenbaum
  1980. ``Stochastic Transition in Two-Dimensional
  {Lennard-Jones} Systems.'' {\em Physics Review A} 22:1709--1719.

\item 
Berti, Massimiliano, Luca Biasco, and Philippe Bolle  2003. ``Drift in Phase Space: a New Variational Mechanism
  with Optimal Diffusion Time.'' {\em Journal de Math\'{e}matiques Pures et
  Appliqu\'{e}s} 82:613--664.

\item 
Bocchieri, P., Antonio Scotti, Bruno Bearzi and A. Loinger 1970. ``Anharmonic Chain With
  {Lennard-Jones} Interaction.'' {\em Physical Review A} 2:213--219.

\item 
Boltzmann, Ludwig 1871. ``Einige allgemeine
  {S\"{a}tze} \"{u}ber {W\"{a}rmegleichgewicht}', {\em Wiener Berichte} 53:670--711.

\item 
Bricmont, Jean 2001. ``Bayes, Boltzmann and
  Bohm: Probabilities in Physics.'' In \emph{Chance in Physics: Foundations and Perspectives}, ed.
     Jean Bricmont, Detlef D\"{u}rr, Maria C. Galavotti, Gian C. Ghirardi, Francesco Pettrucione, and Nino Zanghi, 3--21. Berlin and New York: Springer.

\item 
Callender, Craig 1999. ``Reducing
  Thermodynamics to Statistical Mechanics: The Case of Entropy.'' {\em Journal
  of Philosophy} 96:~348--373.

\item 
--------- 2001. ``Taking Thermodynamics
Too Seriously.'' {\em Studies in History and Philosophy of Modern Physics} 32:539--553.

\item 
Chernov, Nikolai, and Roberto Markarian 2006. {\em Chaotic Billiards}. Providence: American Mathematical Society.

\item 
Chernov, Nikolai, and Lai-Sang Young 2000.
  ``Decay of {Lorentz Gases} and Hard Balls.'' In \emph{Hard
  Ball Systems and the Lorentz Gas}, ed. Domokos Sz\'{a}sz, 89--120.  Berlin: Springer.

\item 
Chirikov, Boris V. 1979. ``A Universal
  Instability of Many-Dimensional Oscillator Systems.'' {\em Physics Reports}
  56:263--379.

\item 
--------- 1991. ``Patterns in Chaos.''
  {\em Chaos, Solitons and Fractals} 1:79--103.

\item 
De~Souza, Vanessa K., and David J. Wales  2005. ``Diagnosing Broken Ergodicity Using an Energy
  Fluctuation Metric.'' {\em The Journal of Chemical Physics} 123:134--504.

\item 
Dellago, Christoph and Harald A. Posch 1997.
  ``Mixing, {Lyapunov} Instability, and the Approach to Equilibrium in a
  Hard-Sphere Gas', {\em Physical Review E} 55:9--12.

\item 
Delshams, Amadeu, and Gemma Huguet 2009.
  ``Geography of Resonances and {Arnold} Diffusion in A Priori Unstable
  {Hamiltonian} Systems.'' {\em Nonlinearity} 22:1997--2077.

\item 
Diana, E., Luigi Galgani, Mario Casartelli, Giulio Casati, and Antonio Scotti 1976. ``Stochastic Transition in a
  Classical Nonlinear Dynamical System: A {Lennard-Jones} Chain.'' {\em
  Theoretical and Mathematical Physics} 29:1022--1027.

\item 
Dizadji-Bahmani, Foad, Roman Frigg, and Stephan Hartmann
  2010. ``Who's Afraid of {Nagelian} Reduction?'' {\em
  Erkenntnis} 73:393-412.

\item 
Donnay, Victor J. 1999. ``Non-Ergodicity of Two
  Particles Interacting Via a Smooth Potential.'' {\em Journal of Statistical
  Physics} 96:1021--1048.

\item 
Donnay, Victor J., and Carlangelo Liverani 1991. ``Potentials on the Two-Torus for Which the {Hamiltonian} Flow is Ergodic.'' {\em Communications in Mathematical Physics} 135:267--302.
\item 
Earman, John, and Mikl\'{o}s R\'{e}dei 1996.
  ``Why Ergodic Theory Does Not Explain the Success of Equilibrium Statistical
  Mechanics.'' {\em The British Journal for the Philosophy of Science} 47:63--78.

\item 
Emch, Gerard G., and Chuang Liu 2002. {\em
  The Logic of Thermostatistical Physics}. Berlin, Heidelberg: Springer.

\item 
Frigg, Roman 2008. ``A Field Guide to Recent
Work on the Foundations of Statistical Mechanics.'' In \emph{The Ashgate Companion to Contemporary Philosophy of Physics}, ed. Dean Rickles, 99--196. London: Ashgate.

\item 
--------- 2009a, ``Probability in {Boltzmannian} Statistical Mechanics.''
In \emph{Time, Chance and Reduction. Philosophical Aspects of
Statistical Mechanics}, ed. Gerhard Ernst, and Andreas H\"{u}ttemann, 92-118. Cambridge: Cambridge University Press.

\item 
--------- 2009b. ``Typicality and the
  Approach to Equilibrium in {Boltzmannian} Statistical Mechanics.'' {\em
  Philosophy of Science (Proceedings)} 76:997--1008.

\item 
--------- 2010. ``Why Typicality Does Not
  Explain the Approach to Equilibrium.'' In \emph{Probabilities, Causes and Propensitites in Physics}, ed. Mauricio  Su\'{a}rez, 77--93. Berlin: Springer.

\item 
Frigg, Roman, and Carl Hoefer 2010. ``Determinism and Chance From a {Humean} Perspective.'' In
\emph{The Present Situation in the Philosophy of Science}, ed. Dennis Dieks,
Wesley Gonzalez, Stephan Harmann, Marcel Weber, Friedrich Stadler, and Thomas Uebel, 351--372.
Berlin and New York: Springer.

\item 
Froeschl\'{e}, Claude, Massimiliano Guzzo, and Elena Lega
  2000. ``Graphical Evolution of the {Arnold} Web: From Order
  to Chaos.'' {\em Science} 289:2108--2110.

\item 
Froeschl\'{e}, Claude, and Jean-Paul Schneidecker
  1975. ``Stochasticity of Dynamical Systems With Increasing
Degrees of Freedom.'' {\em Physical Review A} 12:2137--2143.

\item 
Goldstein, Sheldon 2001. ``{Boltzmann's} Approach
to Statistical Mechanics.'' In \emph{Chance in Physics: Foundations and Perspectives}, ed. Jean Bricmont, Detlef D\"{u}rr, Maria C. Galavotti, Gian C. Ghirardi, Francesco Pettrucione, and Nino Zanghi, 39--54. Berlin and New York: Springer.

\item 
Guzzo, Massimiliano, Elena Lega, and  Claude Froeschl\'{e}
  2005. ``First Numerical Evidence of Global {Arnold}
  Diffusion in Quasi-Integrable Systems.'' {\em Discrete and Continuous Dynamical Systems Series B} 5:687--698.

\item 
Hirsch, Morris W. 1976. {\em Differential
  Topology}. Berlin: Springer.

\item 
Lavis, David 2005. ``Boltzmann and {Gibbs}: An
  Attempted Reconciliation.'' {\em Studies in History and Philosophy of Modern
  Physics} 36:245--273.

\item 
--------- 2011. ``An Objectivist Account of
  Probabilities in Statistical Physics.'' \emph{Probabilities in Physics}, ed. Claus Beisbart, and Stephan Hartmann, forthcoming. Oxford: Oxford University Press, Oxford.

\item 
Markus, Larry and Kenneth R. Meyer 1969.
  ``Generic {Hamiltonians} Are Not Ergodic.'' In \emph{Proceedings of the 9th
  Conference on Nonlinear oscillations}, ed. Yu A. Mitropolsky, 311--332. Kiev: Kiev Naukova Dumka.

\item 
--------- 1974. ``Generic {Hamiltonian} Dynamical Systems Are Neither Integrable Nor Ergodic.'' {\em Memoirs of the American Mathematical Society} 144:1--52.

\item 
Mather, John N.  2004. ``Arnold Diffusion {I}:
  Announcement of Results.'' {\em Journal of Mathematical Sciences}
  124:5275--5289.

\item
McQuarrie, Donald A. 2000. {\em Statistical
  Mechanics}. Sausolito, California: University Science Books.

\item 
Ott, Edward 2002. {\em Chaos in Dynamical
  Systems}. Cambridge: Cambridge University Press.
\item 
Oxtoby, John C.  1980. {\em Measure and
  Category}. New York, Heidelberg, Berlin: Springer.

\item 
Penrose, Oliver 1979. ``Foundations of
  Statistical Physics.'' {\em Reports on Progress in Physics}
  42:1937--2006.

\item 
Penrose, Oliver, and Joel L. Lebowitz 1973. ``Modern Ergodic Theory.'' {\em Physics Today}
  26:23--29.

\item 
Percival, Ian 1986. ``Integrable and
  Nonintegrable {Hamiltonian} Systems.'' In \emph{Nonlinear Dynamics Aspects of Particle
  Accelerators}, ed. John M. Jowett, Melvin Month, and Stuart Turner, 12--36. Berlin, Heidelberg: Springer.

\item 
Pettini, Marco 2007. {\em Geometry and
  Topology in {Hamiltonian} Dynamics and Statistical Mechanics}. New York: Springer.

\item 
Pettini, Macro, and Monica Cerruti-Sola
  1991. ``Strong Stochasticity Thresholds in Nonlinear Large
  {Hamiltonian} Systems: Effect on Mixing Times.'' {\em Physical Review A}
  44:975--987.

\item 
Reichl, Linda E. 1998. {\em A Modern Course in
  Statistical Physics}. New York: Wiley.

\item 
Reidl, Charles J., and Bruce N. Miller 1993. ``Gravity in One Dimension: The Critical Population.'' {\em Physical Review E} 48:4250--4256.

\item 
Sim\'{a}nyi, Nandor 1992. ``The {K}-Property of
  {$N$} Billiard Balls.'' {\em Inventiones Mathematicae} 108:521--548.

\item 
--------- 1999. ``Ergodicity of Hard
  Spheres in a Box.'' {\em Ergodic Theory and Dynamical Systems} 19:741--766.
\item 
--------- 2003. ``Proof of the
  {Boltzmann-Sinai} Ergodic Hypothesis for Typical Hard Disk Systems.'' {\em
  Inventiones Mathematicae} 154:123--178.
\item 
--------- 2009. ``Conditional Proof of
  the {Boltzmann-Sinai} Ergodic Hypothesis.'' {\em Inventiones Mathematicae}
177:381--413.

\item 
Sinai, Yakov G. 1963. ``On the Foundations of the
Ergodic Hypothesis for a Dynamical System of Statistical Mechanics.'' {\em
  Soviet Mathematics Doklady} 4:1818--1822.

\item 
--------- 1970. ``Dynamical Systems With
  Elastic Reflections: Ergodic Properties of Dispersing Billiards.'' {\em
  Uspekhi Matematicheskikh Nauk} 25:141--192.

\item 
Sklar, Lawrence 1993. {\em Physics and Chance:
  Philosophical Issues in the Foundations of Statistical Mechanics}. Cambridge: Cambridge
  University Press.

\item 
Stoddard, Spotswood D., and Joseph Ford. 1973.
  ``Numerical Experiments on the Stochastic Behaviour of a {Lennard-Jones} Gas
  System.'' {\em Physical Review A} 8:1504--1512.
\item 
Sz\'{a}sz, Domokos 1996. ``Boltzmann's Ergodic
  Hypothesis: A Conjecture for Centuries?'' {\em Studia Scientiarum
  Mathematicarum Hungarica}  31:299--322.
\item 
Uffink, Jos 1996. ``Nought But Molecules in
  Motion (Review Essay of {Lawrence Sklar}: {Physics and Chance}).'' {\em
  Studies in History and Philosophy of Modern Physics} 27:373--387.

\item 
--------- 2007. ``Compendium to the
  Foundations of Classical Statistical Physics.'' In \emph{Philosophy of Physics}, ed.
 Jeremy Butterfield and John Earman, 923--1074. Amsterdam: North-Holland.

\item 
Vivaldi, Franco 1984. ``Weak-Instabilities in
  Many-Dimensional {Hamiltonian} Systems.'' {\em Review of Modern Physics}
  56:737--753.

\item 
Vranas, Peter B.M.  1998. ``Epsilon-Ergodicity and
  the Success of Equilibrium Statistical Mechanics.'' {\em Philosophy of
  Science} 65:688--708.

\item 
Werndl, Charlotte 2009a. ``Are Deterministic
  Descriptions and Indeterministic Descriptions Observationally Equivalent?''
  {\em Studies in History and Philosophy of Modern Physics} 40:232--242.

\item 
--------- 2009b. ``Justifying Definitions in Mathematics. Going Beyond Lakatos.''
  {\em Philosophia Mathematica} 17:313--340.

\item 
--------- 2009c. ``What Are the New
  Implications of Chaos for Unpredictability?'' {\em The British Journal for
  the Philosophy of Science} 60:195--220.

\item 
--------- 2011. ``On the Observational
  Equivalence of Continuous-Time Deterministic and Indeterministic
  Descriptions.'' {\em European Journal for the Philosophy of Science}
 DOI: 10.1007/s1319401000115, forthcoming.

\item 
Wright, Harold, and Bruce N. Miller 1984.
  ``Gravity in One Dimension: A Dynamical and Statistical Study.'' {\em Physical
  Review A} 29:1411--1418.

\item 
Yoshimura, Kazuyoshi 1997. ``Strong Stochasticity
  Threshold in Some Anharmonic Lattices.'' {\em Physica D} 104:148--162.

\item 
Zheng, Zhigang, Gang Hu, and Juyuan Zhang
  1996. ``Ergodicity in Hard-Ball Systems and {Boltzmann's}
  Entropy.'' {\em Physical Review E} 53:3246--3253.
  
  \end{list}
  \end{document}